\documentclass{article}
\usepackage{epsfig}
\usepackage{amssymb}
\newcommand{\bfr}{\begin{flushright}}
\newcommand{\efr}{\end{flushright}}
 
\begin{document}
\title{Vacuum polarization near asymptotically anti-de Sitter black holes in odd
dimensions
}
\author{Kiyoshi Shiraishi\\
Akita Junior College, Shimokitade-Sakura,\\
Akita-shi, Akita 010,
Japan\\
and\\
Takuya Maki\\
Department of Physics, Tokyo Metropolitan University, \\
Minami-ohsawa,
Hachioji-shi, Tokyo 192-03, Japan}
\date{Class. Quantum Grav. \textbf{11} (1994) pp.~1687--1696}
\maketitle
\begin{abstract}
Recently, Ba\~nados, Teitelboim and Zanelli obtained
spherically symmetric black hole solutions in a particular class of
Einstein--Lovelock gravity. We derive the propagator in an exact form for
a conformal scalar field in the asymptotically anti-de Sitter black hole
spacetime so as to study the quantum effects of the scalar fields. We
treat the cases in odd dimensions in this paper. We calculate the vacuum
expectation value of $\langle\varphi^2\rangle$ and show its dependence on
the radial coordinate for the five-dimensional case as an example.
\end{abstract}

\section{Introduction}
For a couple of decades quantum field theorists have studied the quantum
field near black holes \cite{1,2,3}. Study of the quantization in the
black hole background is important not only because it is necessary to
attain the complete description of black hole thermodynamics but also
because the quantum back reaction may change the picture of the endpoint
of black hole evaporation.

 Recently, the black hole solution to odd-dimensional Einstein--Lovelock
gravity has been found and the global structure of the spacetime and
thermodynamics at `zero-loop' level have been analysed%
\footnote{Hereafter, we call the black hole solution found by them the `Ba\~nados'
black hole', simply for reasons of brevity.
}\cite{4}. The
black hole solution in odd dimensions approaches anti-de Sitter space in
the asymptotic region and is the higher-dimensional generalization of the
three-dimensional black hole solution \cite{5,6,7,8,9}. It will be
interesting to investigate the nature of the quantum field in the
odd-dimensional black hole background, because in odd dimensions there
is no conformal anomaly at the one-loop level \cite{3}, which is believed
to be closely connected with Hawking radiation and is expected to have
something to do with other quantum effects near the black holes at least
in four dimensions.

The present authors \cite{7} and several groups \cite{8,9} have obtained
the propagator and the vacuum expectation value of
$\langle\varphi^2\rangle$ and stress tensor for a conformally coupled
scalar field in three-dimensional black hole spacetime. In the present
paper, we provide an explicit expression for the scalar field propagator
and the vacuum polarization of $\langle\varphi^2\rangle$ in the
Ba\~nados' black hole background in odd dimensions. The calculation of
$\langle\varphi^2\rangle$ can be performed by much less effort than the
stress tensor. We would regard the calculation of
$\langle\varphi^2\rangle$ as a useful preliminary to the calculation of
the quantum stress tensor.

In section 2, we briefly review the black hole solutions obtained by
Ba\~nados \textit{et al} to make the present paper self-contained. We
obtain the propagator for a conformally coupled massless scalar field in
the black hole spacetime in section 3. The expectation value of
$\langle\varphi^2\rangle$ for the scalar field is derived from the
propagator. In section 4 we compute $\langle\varphi^2\rangle$ in the
five-dimensional black hole spacetime as a concrete example. A summary
is given in section 5.

\section{Black hole solutions in the gravity theory proposed by 
Ba\~nados \textit{et al}}
We consider $D$-dimensional spacetime, where $D$ is assumed to be odd and
written by $D=2m+1$. The action proposed by Ba\~nados \textit{et al},
which contains only two coupling constants, is written by%
\footnote{Please be careful about the notation, which differs slightly
from Ba\~nados et al.
}\cite{14}:
\begin{equation}
I=\kappa_G\sum_{p=0}^m\frac{a^{-D+2p}}{D-2p}\left(
\begin{array}{c}
m \\ p
\end{array}
\right) I_p
\label{2.1}
\end{equation}
with
\begin{equation}
I_p=\int \varepsilon_{a_1\ldots a_D}R^{a_1a_2}\wedge\cdots
\wedge R^{a_{2p-1}a_{2p}}\wedge e^{a_{2p+1}}\wedge\cdots\wedge
e^{a_{D}}\,,
\label{2.2}
\end{equation}
where $\kappa_G$ is a dimensionless constant while $a$ is the coupling
which has the dimension of length in the natural unit system. We take
$\kappa_G=1/(D-2)!A_{D-2}$ with $A_{D-2}=2\pi^{(D-1)/2}\Gamma((D-1)/2)$.
The expression (\ref{2.2}) is often referred to as the dimensionally
continued Euler form.

When there is no matter field coupled to gravity, the equation of
motion is derived from the action as
\begin{equation}
\varepsilon_{a_1\ldots a_D}F^{a_1a_2}\wedge\cdots\wedge
F^{a_{D-2}a_{D-1}}=0\,,
\label{2.3}
\end{equation}
where
\begin{equation}
F^{ab}=R^{ab}+a^{-2}e^a\wedge e^b\,.
\label{2.4}
\end{equation}

Ba\~nados \textit{et al} found the following spherically symmetric
solution: 
\begin{equation}
ds^2=-g^2(r) dt^2+ g^{-2}(r) dr^2+ r^2d\Omega^2\,,
\label{2.5}
\end{equation}
with
\begin{equation}
g^2(r)=1-(1+M)^{1/m}+\left(\frac{r}{a}\right)^2=-\tilde{M}
+\left(\frac{r}{a}\right)^2\,,
\label{2.6}
\end{equation}
where $M$ is the mass of the black hole. We use the parameter $\tilde{M}$
hereafter for convenience. Note that $\tilde{M}=0$ when $M=0$ and
$\tilde{M}$ increases simply with $M$.

For $D=3$, the spacetime is identified as `anti-de Sitter space with
global conical structure'. Therefore we recognize that this spacetime
can be made from the anti-de Sitter space with some identification
procedure \cite{5}. But for $D\ge 5$, one cannot obtain the metric from
the identification process on anti-de Sitter space. In fact, curvature of
the spacetime is no longer constant. This feature is akin to the
difference between cosmic strings and global monopoles, which induce
deficit angles and deficit `solid' angles respectively.

We can rewrite the metric by using a new coordinate:
\begin{equation}
r=r_+ \sec\rho \qquad	(0\le\rho\le\pi/2)	\,,
\label{2.7}
\end{equation}
where $r_+=\sqrt{\tilde{M}}a$.

Then we get
\begin{equation}
ds^2=a^2(\sec\rho)^2(-\kappa^2 \sin^2\rho\, dt^2 +d\rho^2
+\tilde{M}d\Omega^2)\,.
\label{2.8}
\end{equation}
Replacing $t$ with the Euclidean one, we obtain
\begin{equation}
ds_E^2=a^2(\sec\rho)^2(\kappa^2 \sin^2\rho \,d\tau^2 +d\rho^2
+\tilde{M}d\Omega^2)\,,
\label{2.9}
\end{equation}
where $\tau$ has a period $2\pi/\kappa$ with $\kappa=\sqrt{\tilde{M}}/a$.

The non-zero components of the Ricci tensor are:
\begin{eqnarray}
& &R_\tau^\tau=R_\rho^\rho=-\frac{D-1}{a^2}\,,
\label{2.10}
\\
& &R_n^m=\left[
-\frac{D-1}{a^2}+\frac{(D-3)(\tilde{M}+1)}{a^2\tilde{M}}
\cos^2\rho
\right]\delta_n^m\,,
\label{2.11}
\end{eqnarray}
where $m$ and $n$ run over the spherical coordinates. One can see the
fact that the spacetime 
expressed by the metric has constant curvature only if $D=3$. The
scalar curvature is then
\begin{equation}
R=\left[
-\frac{D(D-1)}{a^2}+\frac{(D-2)(D-3)(\tilde{M}+1)}{a^2\tilde{M}}
\cos^2\rho
\right]\,.
\label{2.12}
\end{equation}

In the next section, we construct the propagator in the Euclidean black
hole spacetime in odd dimensions by using the mode sum method.

\section{Two-point function in the asymptotically anti-de Sitter 
black hole spacetime}
We introduce a conformally invariant scalar in the Ba\~nados black hole
background. The wave equation for a conformal scalar is
\begin{equation}
\Box\varphi-\frac{D-2}{4(D-1)}R\varphi=0\,,
\label{3.1}
\end{equation}
where the covariant divergence is defined in terms of the background
metric (\ref{2.9}). The Euclidean propagator $G_H$ satisfies:
\begin{equation}
\left(\Box-\frac{D-2}{4(D-1)}R\right)G_H(x, x')=
-\frac{1}{\sqrt{\det g_{\mu\nu}}}\delta(x, x')\,.
\label{3.2}
\end{equation}

To obtain the propagator, we adopt the mode-sum method. The mode
function obeys the wavefunction. If the function takes the form
\begin{equation}
\varphi_{Nn}(x)=u_{Nn}(\rho)Y_{N\nu}^{(D-2)}(\Omega) e^{in\kappa\tau}\,,
\label{3.3}
\end{equation}
where $Y^{(D-2)}(\Omega)$ is the generalized spherical function \cite{2}
and
$\Omega$ represents the coordinate on a unit $(D 2)$-sphere, the radial
function $u_{Nn}(\rho)$ obeys
\begin{eqnarray}
& &\frac{\cos^{D-2}\rho}{\sin\rho}\frac{d}{d\rho}
\frac{\sin\rho}{\cos^{D-2}\rho}\frac{d}{d\rho}u_{Nn}(\rho)-
\left[\frac{n^2}{\sin^2\rho}+\frac{N(N+D-3)}{\tilde{M}}\right]
u_{Nn}(\rho)\nonumber \\
& &\quad-\frac{D-2}{4(D-1)}\left[-\frac{D(D-1)}{\cos^2\rho}+
\frac{(\tilde{M}+1)(D-2)(D-3)}{\tilde{M}}\right] u_{Nn}(\rho)=0\,. 
\label{3.4}
\end{eqnarray}

The general solution for this differential equation is given by a
linear combination of the two independent functions:
\begin{equation}
u_{Nn}(\rho)=(\cos\rho)^{(D-2)/2}[\alpha P^n_{-1/2\pm i\mu}(\cos\rho)
+\beta Q^n_{-1/2\pm i\mu}(\cos\rho)]\,,	
\label{3.5}
\end{equation}
where $P^n_\nu(x)$ and $Q^n_{\nu}(x)$ are the Legendre functions and
\begin{equation}
\mu=\sqrt{\frac{1}{\tilde{M}}\left[
\left(N+\frac{D-3}{2}\right)^2+\frac{(\tilde{M}+ 1)(D-3)}{4(D-1)}\right]
}\,.
\label{3.6}
\end{equation}

Instead of (\ref{3.5}), we will choose a set of real functions $(\cos
\rho)^{(D-2)/2}$ $\{P^n_{-1/2+i\mu}(\cos\rho),$ $
P^n_{-1/2+i\mu}(-\cos\rho)\}$ as mode functions.

We can construct the Euclidean propagator from the mode functions. In
general, the mode sum takes the following form:
\begin{eqnarray}
& &G_H(\rho, \tau, \Omega; \rho', \tau', \Omega')\nonumber \\
& &=\frac{\kappa}{2\pi}\sum_{n=-\infty}^\infty e^{in\kappa(\tau-\tau')}
\sum_{N=0}^\infty\sum_\nu	Y^{(D-2)}_{N\nu}(\Omega)
Y^{(D-2)*}_{N\nu}(\Omega')f(\cos\rho_{<}) g(\cos\rho_{>})\,,
\label{3.7}
\end{eqnarray}
where $\rho_{<}<\rho_{>}$, and, $f$ and $g$ are solutions of
equation (\ref{3.4}), chosen to reflect boundary conditions.

At $\rho=0$, which corresponds to the horizon,
$P^n_{-1/2+i\mu}(\cos\rho)$ has a finite value and is regular while
$P^n_{-1/2+i\mu}(-\cos\rho)$ and $Q^n_{-1/2+i\mu}(\cos\rho)$ not. Thus we
must take
$f\approx(\cos\rho)^{(D-2)/2}P^n_{-1/2+i\mu}(\cos\rho)$. At
$\rho=-\pi/2$, which corresponds to spatial infinity, various boundary
conditions can be considered, because anti-de Sitter space is not
globally hyperbolic \cite{15}. We assume
\begin{equation}
g(\cos\rho)\approx(\cos\rho)^{(D-2)/2}[P^n_{-1/2+i\mu}(-\cos\rho)-
\alpha
P^n_{-1/2+i\mu}(\cos\rho)]	
\label{3.8}
\end{equation}
where $\alpha$ is constant. $\alpha=1(-1)$ corresponds to Dirichlet
(Neumann) boundary condition at $\rho=\pi/2$. The condition $\alpha=0$
is called the transparent boundary condition, according to \cite{15}.

The normalizations of the mode functions are determined by the
Wronskian condition on the two functions. Using the formula
\begin{equation}
\left|\begin{array}{cc}
P^n_{-1/2+i\mu}(\cos\rho)	& \frac{d}{d\rho}P^n_{-1/2+i\mu}(\cos\rho)\\
P^n_{-1/2+i\mu}(-\cos\rho) &  \frac{d}{d\rho}P^n_{-1/2+i\mu}(-\cos\rho)
\end{array}\right|=-\frac{2}{\pi}\cosh\mu\pi\frac{(-1)^n}{\sin\rho}\,,
\label{3.9}
\end{equation}
we get the expression for the propagator:	
\begin{eqnarray}
& &G_H(\rho, \tau, \Omega; \rho', \tau', \Omega')
=\frac{\kappa}{2\pi}\sum_{n=-\infty}^\infty e^{in\kappa(\tau-\tau')}
\sum_{N=0}^\infty\sum_\nu	Y^{(D-2)}_{N\nu}(\Omega)
Y^{(D-2)*}_{N\nu}(\Omega')
\nonumber \\
& &\qquad\times (\cos\rho\cos\rho')^{(D-2)/2}\frac{1}{\kappa r_+^{D-2}}
\frac{\pi}{2}\frac{1}{\cosh\mu\pi}	P^n_{-1/2+i\mu}(\cos\rho_{<})
\nonumber \\
& &\qquad\times
[P^{-n}_{-1/2+i\mu}(-\cos\rho_{>})-\alpha
P^{-n}_{-1/2+i\mu}(\cos\rho_{>})]\,.
\label{3.10}
\end{eqnarray}
Applying the addition theorem \cite{10} to this then we get
\begin{eqnarray}
& &G_H(\rho, \tau, \Omega; \rho', \tau', \Omega')
=\frac{(\cos\rho\cos\rho')^{(D-2)/2}}{2\pi r_+^{D-2}}
\sum_{N=0}^\infty\sum_\nu	Y^{(D-2)}_{N\nu}(\Omega)
Y^{(D-2)*}_{N\nu}(\Omega')
\nonumber \\
& &\qquad\times 
\frac{\pi}{2}\frac{1}{\cosh\mu\pi}[P_{-1/2+i\mu}(-\cos\rho
\cos\rho'-\sin\rho\sin\rho'\cos\kappa(\tau-\tau'))
\nonumber \\
& &\qquad \qquad \qquad \qquad
-\alpha P_{-1/2+i\mu}(\cos\rho
\cos\rho'-\sin\rho\sin\rho'\cos\kappa(\tau-\tau'))]\,.
\label{3.11}
\end{eqnarray}

Using the integral representation for the Legendre functions \cite{10},
we obtain
\begin{eqnarray}
& &G_H(\rho, \tau, \Omega; \rho', \tau', \Omega')
=\frac{(\cos\rho\cos\rho')^{(D-2)/2}}{2\pi r_+^{D-2}}
\sum_{N=0}^\infty\sum_\nu	Y^{(D-2)}_{N\nu}(\Omega)
Y^{(D-2)*}_{N\nu}(\Omega')
\nonumber \\
& &\qquad\times 
\left[\int_0^\infty \frac{\cos\mu\phi\, d\phi}{\sqrt{2(\cosh\phi-\cos\rho
\cos\rho'-\sin\rho\sin\rho'\cos\kappa(\tau-\tau'))}}\right.
\nonumber \\
& &\qquad \qquad \left.
-\alpha \int_0^\infty \frac{\cos\mu\phi\,
d\phi}{\sqrt{2(\cosh\phi+\cos\rho
\cos\rho'-\sin\rho\sin\rho'\cos\kappa(\tau-\tau'))}}\right]\,,
\label{3.12}
\end{eqnarray}
where $\mu$ is given by (\ref{3.6}).

For $D=3$, this representation is further simplified to (with noting
$\tilde{M}=M$ in this case)
\begin{eqnarray}
& &G_H(\rho, \tau, \theta; \rho', \tau', \theta')
\nonumber \\
& &=\sum_{k=-\infty}^\infty
\frac{(\cos\rho)^{1/2}(\cos\rho')^{1/2}}{4\sqrt{2}\pi
a\sqrt{\cosh\sqrt{M}(\theta-\theta'+2\pi k)-\cos\rho
\cos\rho'-\sin\rho\sin\rho'\cos\kappa(\tau-\tau')}}
\nonumber \\
& & 
-\alpha
\sum_{k=-\infty}^\infty\frac{(\cos\rho)^{1/2}(\cos\rho')^{1/2}}{4\sqrt{2}\pi
a\sqrt{\cosh\sqrt{M}(\theta-\theta'+2\pi k)+\cos\rho
\cos\rho'-\sin\rho\sin\rho'\cos\kappa(\tau-\tau')}}\,.
\label{3.13}
\end{eqnarray}

This coincides with the propagator constructed from that in the
three-dimensional anti-de Sitter space with identification process.

In the next section, we compute the vacuum value for
$\langle\varphi^2\rangle$ by using the Euclidean propagator (\ref{3.11})
or (\ref{3.12}).

\section{Calculation of $\langle\varphi^2\rangle$}
It is easy to calculate $\langle\varphi^2\rangle$ from the Euclidean
propagator. The vacuum value $\langle\varphi^2\rangle$ is defined as
\cite{2}
\begin{equation}
\langle\varphi^2\rangle(x)=\lim_{x'\rightarrow x}
 (G_H(x, x')-G^{div}_H(x, x'))\,,
\label{4.1}
\end{equation}
where $G^{div}_H$ denotes the divergent part in the Euclidean
propagator. In the spherical black hole background,
$\langle\varphi^2\rangle$ is given as a function of $\rho$.

We assume the two points $x$ and $x'$ have common values of the
coordinates $\tau=\tau'$ and $\Omega=\Omega'$: In this case $G_H$
becomes a function of $\rho$ and $\rho'$.

According to \cite{2}, the sum of the spherical functions for
$\Omega=\Omega'$ can be written as
\begin{equation}
\sum_\nu Y_{N\nu}^{(D-2)}(\Omega)Y_{N\nu}^{(D-2)*}(\Omega)=
\frac{(N+m-1)\Gamma(N+2m-2)\Gamma(m-1)}{2\pi^m\Gamma(N+1)\Gamma(2m-2)}\,,
\label{4.2}
\end{equation}
where $m=(D-1)/2$. Using this formula, we find
\begin{eqnarray}
& &G_H(\rho, \rho')=\frac{(\cos\rho\cos\rho')^{m-1/2}}{4\sqrt{2}
\pi^{m+1}r_+^{2m-1}}\frac{\Gamma(m-1)}{\Gamma(2m-2)}\sum_{N=0}^\infty
\frac{(N+m-1)\Gamma(N+2m-2)}{\Gamma(N+1)}\nonumber \\
& &\qquad\times\left[\int_0^\infty
\frac{\cos\mu\phi\,d\phi}{\sqrt{\cosh\phi-\cos(\rho-\rho')}}
-\alpha\int_0^\infty
\frac{\cos\mu\phi\,d\phi}{\sqrt{\cosh\phi+\cos(\rho-\rho')}}
\right]\,,
\label{4.3}
\end{eqnarray}
where $\mu$ is given by (3.6)

The denominator of the integrand can be expanded by using the Legendre
function:
\begin{equation}
\frac{1}{\sqrt{2(\cosh\phi-\cos\beta)}}=\sum_{q=0}^\infty
P_q(\cos\beta)\,e^{-(q+1/2)\phi}\,.
\label{4.4}
\end{equation}
Using this expansion, we carry out the integration over $\phi$ and get:
\begin{eqnarray}
& &G_H(\rho, \rho')=\frac{(\cos\rho\cos\rho')^{m-1/2}}{4
\pi^{m+1}r_+^{2m-1}}\frac{\Gamma(m-1)}{\Gamma(2m-2)}\sum_{N=0}^\infty
\frac{(N+m-1)\Gamma(N+2m-2)}{\Gamma(N+1)}\nonumber \\
& &\quad\times\left[\sum_{q=0}^\infty\int_0^\infty
(P_q(\cos(\rho-\rho'))-\alpha P_q(-\cos(\rho+\rho')))
e^{-(q+1/2)\phi}\cos\mu\phi\,d\phi
\right]\nonumber \\
& &=\frac{(\cos\rho\cos\rho')^{m-1/2}}{4
\pi^{m+1}r_+^{2m-1}}\frac{\Gamma(m-1)}{\Gamma(2m-2)}\sum_{N=0}^\infty
\frac{(N+m-1)\Gamma(N+2m-2)}{\Gamma(N+1)}\nonumber \\
& &\quad\times\left[\sum_{q=0}^\infty
\frac{q+1/2}{(q+1/2)^2+\mu^2}
(P_q(\cos(\rho-\rho'))-\alpha P_q(-\cos(\rho+\rho')))
\right]\,.
\label{4.5}
\end{eqnarray}

For odd-dimensional spacetime, the summation on $N$ can be carried out by
the technique introduced by \cite{11} and \cite{12}.

The expression for $G^{div}$, on the other hand, can be found by the
method of De Witt \cite{13} and Christensen \cite{14} (and developed by
many other authors). The analysis of the
point-splitting method is rather simple in our case, because the line
element in the radial direction is the same as that in Euclidean anti-de
Sitter space.

Now we show the calculation of $\langle\varphi^2\rangle$ for the
five-dimensional case ($D=5$, $m=2$) as a concrete example. In the
higher-dimensional cases, the calculation is tedious but straightforward.

First we rewrite equation (\ref{4.5}) by adding a parameter integration,
in order to handle the summation over $N$. We consider the following
expression:
\begin{eqnarray}
& &\!\!\!\!G_H(\rho,\rho')=\frac{(\cos\rho\cos\rho')^{3/2}}{4\pi^3r_+^3}
\sum_{q=0}^\infty\left(q+\frac{1}{2}\right)
\left(P_q(\cos(\rho-\rho'))-\alpha P_q(-\cos(\rho+\rho'))\right)
\nonumber \\
& &\times\left[\sum_{N=0}^\infty(N+1)^2\int_0^\infty
	\exp\left\{-\left[\left(q+\frac{1}{2}\right)^2+
\frac{(N+1)^2}{\tilde{M}}+\frac{\tilde{M}+1}{8\tilde{M}}
\right]t\right\}dt\right]\,.
\label{4.6}
\end{eqnarray}

We apply Poisson's summation formula to the sum over $N$. Then we get
\begin{eqnarray}
& &\!\!\!\!G_H(\rho,\rho')=\frac{(\cos\rho\cos\rho')^{3/2}}{8\pi^3a^3}
\sqrt{\pi}
\sum_{q=0}^\infty\left(q+\frac{1}{2}\right)
\left(P_q(\cos(\rho-\rho'))-\alpha P_q(-\cos(\rho+\rho'))\right)
\nonumber \\
& &\times\sum_{L=0}^\infty\int_0^\infty
\frac{dt}{t^{3/2}}\left(\frac{1}{2}-\frac{\pi^2\tilde{M}}{t}L^2\right)
\exp\left[-\left(q+\frac{1}{2}\right)^2t
-\frac{\tilde{M}+1}{8\tilde{M}}t-\frac{\pi^2\tilde{M}}{t}L^2
\right]\,.
\label{4.7}
\end{eqnarray}

In this expression (4.7), we find that the divergent contributions when
$\rho\rightarrow\rho'$ are 
contained in the $L=0$ term in the sum.

Taking care of this fact, we integrate the expression over $t$ and
obtain:
\begin{eqnarray}
& &G_H(\rho, \rho')=\frac{(\cos\rho\cos\rho')^{3/2}}{8\pi^2a^3}
\Bigl\{\frac{1}{[2(1-\cos(\rho-\rho'))]^{3/2}}-\alpha\frac{1}{[2(1+\cos(\rho+
\rho'))]^{3/2}}\nonumber \\
& &-\frac{\tilde{M}+1}{16\tilde{M}}\left[
\frac{1}{\sqrt{2(1-\cos(\rho-\rho'))}}-\alpha\frac{1}{\sqrt{2(1+\cos(\rho+
\rho'))}}
\right]\Bigr\}\nonumber \\
& &+\frac{(\cos\rho\cos\rho')^{3/2}}{8\pi^2a^3}
\frac{1}{\sqrt{4\pi}}
\sum_{q=0}^\infty
\left(P_q(\cos(\rho-\rho'))-\alpha P_q(-\cos(\rho+\rho'))\right)
\nonumber \\
&
&\times\left[\sum_{n=2}^\infty\frac{(-1)^n}{n!}
\frac{\Gamma(n-1/2)}{(q+1/2)^{2n-2}}\left(
\frac{\tilde{M}+1}{8\tilde{M}}
\right)^{n}\right]-\frac{(\cos\rho\cos\rho')^{3/2}}{8\pi^2a^3}2\nonumber
\\
& &\times\sum_{q=0}^\infty\left(q+\frac{1}{2}\right)\sqrt{
\left(q+\frac{1}{2}\right)^2+\frac{\tilde{M}+1}{8\tilde{M}}}\Bigl(
P_q(\cos(\rho-\rho'))\nonumber \\
& &-\alpha P_q(-\cos(\rho+\rho'))\Bigr)
\frac{1}{\exp\left(
2\pi\sqrt{\tilde{M}}\sqrt{
\left(q+\frac{1}{2}\right)^2+\frac{\tilde{M}+1}{8\tilde{M}}}
\right)-1}\,.
\label{4.8}
\end{eqnarray}

On the other hand, $G^{div}$ can be obtained by the point-splitting
method
\cite{13,14}. In the present case, we take
\begin{equation}
G^{div}_H=\frac{\Delta^{1/2}}{8\pi^2}	
\frac{1}{(2\sigma)^{3/2}}
\left\{1+\frac{a_1}{2}(2\sigma)+\cdots
\right\}\,,
\label{4.9}
\end{equation}
where
\begin{eqnarray}
\Delta^{1/2}&=&1+\frac{1}{12}R_{\mu\nu}\sigma^\mu\sigma^\nu-
\frac{1}{24}R_{\mu\nu;\lambda}\sigma^\mu\sigma^\nu\sigma^\lambda+
\cdots\,,
\label{4.10}\\
a_1&=&\left(\frac{1}{6}-\frac{3}{16}\right)R\,,
\label{4.11}
\end{eqnarray}
where
$\sigma$ is the geodesic distance and $\sigma_\mu=\sigma_{,\mu}$. Now,
since we consider the radial separation, we take
\begin{equation}
\sigma=\frac{a^2}{2}\left(\cosh^{-1}\frac{1-\sin\rho\sin\rho'}
{\cos\rho\cos\rho'}	\right)^2\,.
\label{4.12}
\end{equation}

Substituting (\ref{4.10})-(\ref{4.12}) into (\ref{4.9}), we find
$G^{div}$ in the limit of small separation, $\Delta\rho=\rho-\rho'$:
\begin{equation}
G^{div}(\rho, \rho')=\frac{1}{8\pi^2a^3}
\left\{\frac{\cos^3\rho}{\Delta\rho^3}+\frac{\cos\rho}{\Delta\rho}
\left[-\frac{3}{8}+\frac{\tilde{M}-1}{16\tilde{M}}\cos^2\rho\right]
+O(\Delta\rho)
\right\}\,.
\label{4.13}
\end{equation}
This behaviour for small $\Delta\rho$ turns out to be the same as the
small $\Delta\rho$ limit of
\begin{equation}
\frac{(\cos\rho\cos\rho')^{3/2}}{8\pi^2a^3}
\left\{
\frac{1}{[2(1-\cos(\rho-\rho'))]^{3/2}}-\frac{\tilde{M}+1}{16\tilde{M}}
\frac{1}{\sqrt{2(1-\cos(\rho-\rho'))}}
\right\}\,,
\label{4.13}
\end{equation}
which appears in the two-point function in the five-dimensional anti-de
Sitter black hole background (\ref{4.8}).

Consequently, we obtain the expression for the vacuum expectation value
$\langle\varphi^2\rangle$ around the Ba\~nados' black hole in five
dimensions:
\begin{eqnarray}
&
&\langle\varphi^2\rangle(\rho)=\frac{(\cos\rho)^{3}}{8\pi^2a^3}\alpha
\left[-\frac{1}{[2(1+\cos(2\rho))]^{3/2}}+
\frac{\tilde{M}+1}{16\tilde{M}}
\frac{1}{\sqrt{2(1+\cos(2\rho))}}\right]\nonumber \\
& &+\frac{(\cos\rho)^{3}}{8\pi^2a^3}
\frac{1}{\sqrt{4\pi}}
\sum_{q=0}^\infty
\left(1-\alpha P_q(-\cos(2\rho))\right)
\nonumber \\
&
&\times\left[\sum_{n=2}^\infty\frac{(-1)^n}{n!}
\frac{\Gamma(n-1/2)}{(q+1/2)^{2n-2}}\left(
\frac{\tilde{M}+1}{8\tilde{M}}
\right)^{n}\right]\nonumber
\\
&
&-\frac{(\cos\rho)^{3}}{8\pi^2a^3}2
\sum_{q=0}^\infty\left(q+\frac{1}{2}\right)\sqrt{
\left(q+\frac{1}{2}\right)^2+\frac{\tilde{M}+1}{8\tilde{M}}}\Bigl(
1-\alpha P_q(-\cos(2\rho))\Bigr)\nonumber \\
& &
\times\frac{1}{\exp\left(
2\pi\sqrt{\tilde{M}}\sqrt{
\left(q+\frac{1}{2}\right)^2+\frac{\tilde{M}+1}{8\tilde{M}}}
\right)-1}\,.
\label{4.15}
\end{eqnarray}
 
\begin{figure}[ht]
\begin{center}
\includegraphics[width=5cm]{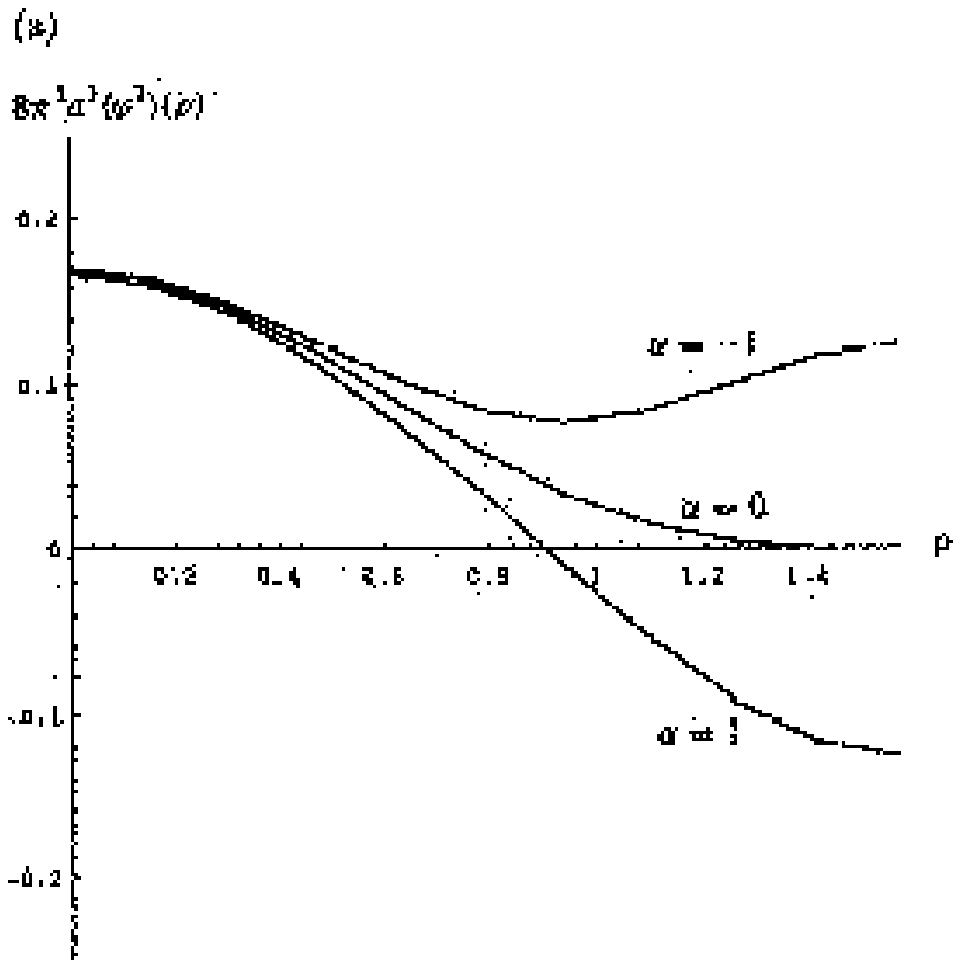}\\
\centering{(a)}\\
\includegraphics[width=5cm]{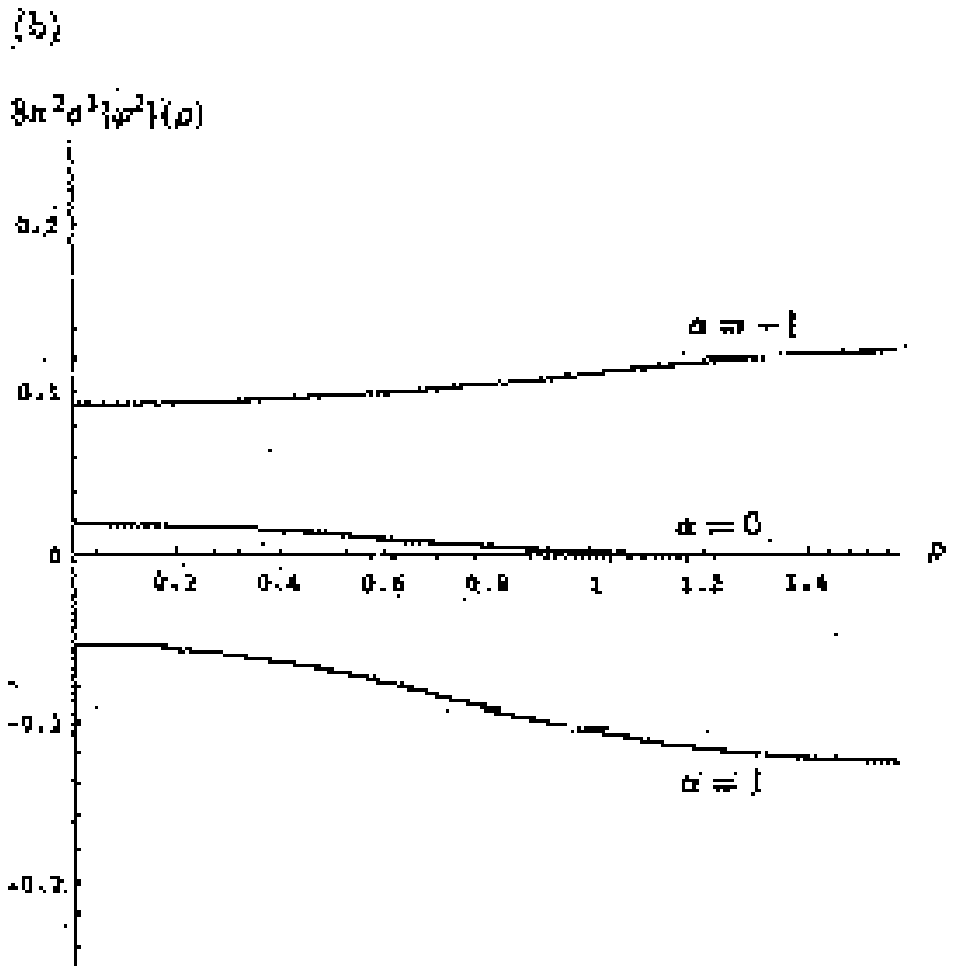}\\
\centering{(b)}
\caption{The magnitude of the vacuum polarization $8\pi^2 a^3
\langle\varphi^2\rangle(\rho)$ as a function of $\rho$ when (a)
$\tilde{M}=0.1$ and (b) $\tilde{M}=1$. The curves correspond to
$\alpha=-1, 0, 1$ as indicated.}
\label{f1}
\end{center}
\end{figure}

For $\alpha=0$ (the transparent boundary condition), the dependence on
$\rho$ becomes very simple. In terms of the original coordinate $r$ in
(\ref{2.5}), $\langle\varphi^2\rangle$ is proportional to $r^{-3}$.
Therefore using this original coordinate, we can consider continuation
of the result to the inside of the black hole. The value of
$\langle\varphi^2\rangle$ diverges only at the origin $r=0$ (for any
$\alpha$).

On the other hand, $\langle\varphi^2\rangle$ approaches zero in the
limit of $r\rightarrow\infty$ if and only if $\alpha=0$. The value of
$\langle\varphi^2\rangle$ at spatial infinity is a constant,
$-\alpha/(64\pi^2a^3)$. This is the value for $\langle\varphi^2\rangle$
in the exact five-dimensional anti-de Sitter space. The numerical
evaluation of $8\pi^2 a^3\langle\varphi^2\rangle(\rho)$ is plotted in
figure \ref{f1} for $\alpha=-1, 0$ and $1$ when $\tilde{M}=0.1$ and
$\tilde{M}=1$.

\begin{figure}[ht]
\begin{center}
\includegraphics[width=5cm]{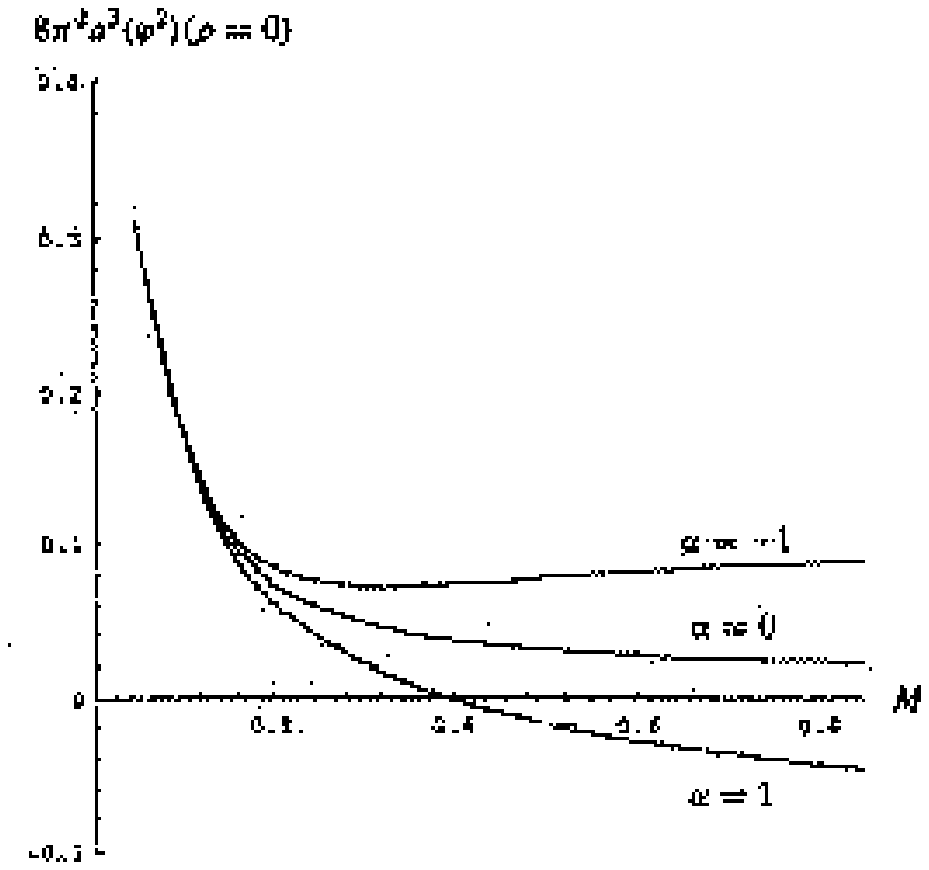}
\caption{The dependence of the vacuum polarization $8\pi^2 a^3
\langle\varphi^2\rangle$ at the horizon as a function of $\tilde{M}$.
The curves correspond to
$\alpha=-1, 0, 1$ as indicated.}
\label{f2}
\end{center}
\end{figure}

The dependence of $\langle\varphi^2\rangle$ at the black hole horizon
($\rho=0; r=r_+$) on $\tilde{M}$ is shown in figure \ref{f2}. The value
of $\langle\varphi^2\rangle$ at the horizon approaches a constant, which
is approximately $1/(8\pi^2a^3) (0.0081 - 0.099\alpha)$, in the large
$\tilde{M}$ limit. We may remark that this depends only on the length
scale $a$.

\section{Summary and prospects}
We have obtained a representation for the Euclidean propagator for a
conformal scalar field in the asymptotically anti-de Sitter black hole
spacetime, which has been found by Ba\~nados et al.

Using the exact propagator, we have computed the vacuum expectation
value $\langle\varphi^2\rangle$ for a conformally coupled massless
scalar field in the five-dimensional case. The value of
$\langle\varphi^2\rangle$ is not positive definite in general.

For large values of the black hole mass, the quantum fluctuation at the
black hole horizon does not vanish in the five-dimensional case even if
a = 0, and perhaps in general higher dimensions. This shows a contrast
to the three-dimensional case, where the amount of the fluctuation
approaches the value of that in the exact anti-de Sitter space.

The amount of quantum fluctuation in the odd-dimensional black hole
spacetime cannot be expressed by simple analytic functions of the
parameter such as the black hole mass. This is a general feature of
field theory in odd dimensions \cite{11,12}.

Using the propagator, we should compute the expectation value for the
stress tensor of quantum fields near the Ba\~nados black holes as a next
step. Then the black hole thermodynamics including quantum fluctuation
of the field could be studied.


\end{document}